\documentclass[a4paper]{iopart}
\pdfoutput=1
\usepackage[utf8]{inputenc}
\usepackage{amsfonts,amsmath,amssymb,graphicx}

\usepackage{xcolor}
\usepackage[square,numbers]{natbib}

\usepackage[colorlinks]{hyperref}
\definecolor{darkred}{rgb}{0.5,0,0}
\definecolor{darkgreen}{rgb}{0,0.5,0}
\definecolor{darkblue}{rgb}{0,0,0.5}
\hypersetup{colorlinks,
linkcolor=darkred,
filecolor=darkgreen,
urlcolor=darkblue,
citecolor=darkgreen}

\usepackage{orcidlink}

\usepackage{braket}

\newcommand{\ud}{\mathrm{d}}
\newcommand{\nep}{\operatorname{e}}

\DeclareMathOperator\sign{sign}

\begin{document}

\title[Classical and quantum chaotic synchronization]{Classical and quantum chaotic synchronization in coupled dissipative time crystals}

\author{Eli\v ska Postavová~\orcidlink{0009-0001-7841-8824}}
\address{Department of Optics, Faculty of Science, Palacký University, 17. listopadu 12, 77146 Olomouc, Czech Republic}

\author{Gianluca Passarelli~\orcidlink{0000-0002-3292-0034}}
\address{Dipartimento di Fisica ``E. Pancini'', Università di Napoli Federico II, I-80126 Napoli, Italy}

\author{Procolo Lucignano~\orcidlink{0000-0003-2784-8485}}
\address{Dipartimento di Fisica ``E. Pancini'', Università di Napoli Federico II, I-80126 Napoli, Italy}

\author{Angelo Russomanno~\orcidlink{0009-0000-1923-370X}}
\address{Dipartimento di Fisica ``E. Pancini'', Università di Napoli Federico II, I-80126 Napoli, Italy~\footnote{Corresponding author, email angelo.russomanno@unina.it}}

\begin{abstract}
We investigate the dynamics of two coherently coupled dissipative time crystals. In the classical mean-field limit of infinite spin length, we identify a regime of chaotic synchronization, marked by a positive largest Lyapunov exponent and a Pearson correlation coefficient close to one. At the boundary of this regime, the Pearson coefficient varies abruptly, marking a crossover between staggered and uniform 
$z$-magnetization. To address finite-size quantum dynamics, we employ a quantum-trajectory approach and study the trajectory-resolved expectations of subsystem $z$-magnetizations. Their histograms over time and trajectory realizations exhibit maxima that undergo a staggered-to-uniform crossover analogous to the classical one. In analogy with the classical case, we interpret this behavior as quantum chaotic synchronization, with dissipative quantum chaos highlighted by the steady-state density matrix exhibiting Gaussian Unitary Ensemble statistics. The classical and quantum crossover points are different due to the noncommutativity of the infinite-time and infinite-spin-magnitude limits and the role played by entanglement in the quantum case, quantified via the two-subsystem entanglement entropy.
\end{abstract}

\section{Introduction}

Synchronization of weakly coupled nonlinear self-sustained oscillators is a ubiquitous phenomenon in nature, observed in systems ranging from coupled neon tubes, to heart cells, to people applauding in a theater~\cite{ROS01a}. An especially important case is the synchronization of many oscillators with slightly different frequencies, elegantly captured by the Kuramoto model~\cite{Kuramoto:1975ebm,Kuramoto2003}. Recent progress in experimental control has brought quantum fluctuations to the forefront. Since the pioneering work of~\cite{PhysRevLett.111.103605}, quantum synchronization has been extensively studied in a variety of bipartite open quantum systems~\cite{Giorgi_2013,PhysRevA.95.043807,SciPostPhys.12.3.097,PhysRevA.109.033718,Murtadho_2023,Solanki_2023,Cabot_2021,Shen_2023,Lee_2014,Walter_2014,li2025twobodydissipatorengineeringenvironmentinduced,PhysRevLett.111.234101}, as well as in many-body settings where collective quantum synchronization emerges~\cite{PhysRevLett.131.190402,Cabot_2019,W_chtler_2023,W_chtler_2024,PhysRevX.15.011010}, even showing robustness of Kuramoto synchronization to quantum noise~\cite{PhysRevA.108.032219}. Experimental realizations have demonstrated entrainment to external drives using cold atoms~\cite{Laskar_2020}, trapped ions~\cite{PhysRevResearch.5.033209,li2025experimentalrealizationsynchronizationquantum}, nuclear spins~\cite{Krithika_2022}, and superconducting circuits~\cite{Koppenh_fer_2020}. In the classical domain, synchronization can also occur between chaotic oscillators~\cite{strogatz:2000,PhysRevLett.64.821,ROS01a,pecora_carro,PhysRevLett.71.65}, a phenomenon with no established quantum counterpart.

A related concept is that of time crystals, where a many-body system exhibits persistent oscillations in an order parameter, thus breaking continuous time-translation symmetry~\cite{Sacha_2017,RevModPhys.95.031001}. In driven, unitary systems, this behavior appears as a collective subharmonic response~\cite{PhysRevLett.117.090402,PhysRevLett.116.250401}, while in dissipative systems it emerges as spontaneous oscillations of collective magnetization in the thermodynamic limit~\cite{PhysRevLett.121.035301}. In this limit, such behavior often reduces to mean-field nonlinear dynamics with self-sustained oscillations. This naturally raises the question of whether coupled time crystals can synchronize. Following the first investigation of synchronization in many dissipative time crystals~\cite{PhysRevLett.128.080603}, subsequent works refined this picture~\cite{PhysRevLett.133.260403}, culminating in a recent study of synchronization between two dissipative (or boundary) time crystals in the thermodynamic limit~\cite{solanki2024chaostimedissipativecontinuous}.  These results revealed distinct synchronized phases, with oscillations that can be either in phase or out of phase. (The dynamics of two coupled boundary time crystals has also been considered from a different perspective in~\cite{PhysRevB.103.184308,paulino2025thermodynamicscoupledtimecrystals}.)

In this contribution we work with the same model, focusing on signatures of chaos and synchronization in both classical and quantum regimes. In the mean-field infinite-spin limit, we characterize dynamics using the largest Lyapunov exponent (LLE) and the Pearson coefficient computed over long times and random initial conditions. We identify the onset of chaotic synchronization: as the coupling strength increases, the Pearson coefficient jumps toward a positive value near one in correspondence with the LLE becoming positive, indicating simultaneous onset of chaos and enhancement of synchronization, akin to chaotic synchronization of Lorenz systems~\cite{pecora_carro,PhysRevLett.71.65}. Alongside, the time-averaged $z$-magnetizations exhibit a crossover from staggered to uniform values across the two subsystems.

To probe quantum effects at finite spin magnitude, we employ the quantum-trajectory method and analyze trajectory-resolved 
$z$-magnetizations. Evaluating for each subsystem the histogram of the $z$-magnetization expectations, over time and the ensemble of trajectory realizations, one can see a maximum that displays a sharp crossover from a staggered to a uniform regime across the two subsystems. This behavior closely echoes the classical case and suggests the emergence of quantum chaotic synchronization. This interpretation is strengthened by the steady-state density matrix, whose spectrum exhibits Gaussian unitary ensemble statistics that marks quantum chaos, and by the behavior of the bipartite entanglement entropy, highlighting the central role of entanglement in quantum synchronization~\cite{Lee_2014,PhysRevLett.111.103605,Giorgi_2013,PhysRevLett.132.183803}. Notably, the crossover points in the quantum and classical cases do not coincide, due to the noncommutativity of the infinite-time and infinite-spin limits: While the classical system exhibits persistent oscillations, the quantum dynamics relaxes to a nonequilibrium steady state~\cite{Delmonte_2025}. Nevertheless, both regimes display a qualitatively similar synchronization crossover, although the physics is different and in the quantum case  a role in the crossover is played by entanglement.

The paper is organized as follows. Section~\ref{mod:sec} introduces the model in both the classical mean-field and quantum finite-size settings. Section~\ref{dim:sec} examines classical chaotic synchronization, focusing on Lyapunov exponents, Pearson coefficients, and magnetization regimes. Section~\ref{qua:sec} presents the finite-spin quantum dynamics, highlighting crossover behavior and the role of entanglement. Section~\ref{con:sec} concludes with a discussion of the relation between classical and quantum chaotic synchronization.

\section{Model}\label{mod:sec}

\subsection{Lindblad equation and classical mean field limit}

We consider a system consisting of two coherently coupled spins of magnitude $S$, evolving under the Lindblad dynamics introduced in Ref.~\cite{solanki2024chaostimedissipativecontinuous}. The unitary part of the evolution is provided by the following Hamiltonian (we will work in natural units, where $\hbar = 1$),
\begin{equation}\label{ham:eqn}
  \hat{H} = \Omega\hat{S}_A^x+\Omega\hat{S}_B^x+ \frac{\Gamma}{2S}\left(\hat{S}_A^+\hat{S}_B^{-}+\hat{S}_A^-\hat{S}_B^{+}\right)\,,
\end{equation}
where $\hat{S}_\alpha^j$ can be considered as the sum of $N$ spin-$1/2$ degrees of freedom, $\hat{S}_\alpha^j = \frac12\sum_{l=1}^N\hat{\sigma}_{j\,l}^\alpha$ (with $\alpha \in \lbrace x, y, z \rbrace$ and $j \in \lbrace A, B \rbrace$). Moreover, $\hat{S}_j^\pm = \hat{S}_j^x \pm i \hat{S}_j^y$. By choosing $S=N/2$, the system is restricted to the Hilbert subspace with maximum total spin of the $N$ spin-$1/2$ degrees of freedom. Therefore, the limit $S\to\infty$ corresponds to a thermodynamic limit, as shown in Ref.~\cite{PhysRevLett.121.035301}. The parameter $\Gamma$ determines the strength of the coherent coupling between the subsystems and is renormalized by the factor $1/S$ to ensure the correct, extensive scaling of the interaction term in the thermodynamic limit. The parameter $\Omega$ is the driving frequency of the two subsystems when uncoupled ($\Gamma = 0$). 

Additionally, the evolution includes a dissipative component described by the Lindblad equation
\begin{equation}
   \frac{\ud \hat{\rho}_t}{\ud t} = -i \left[\hat{H}, \hat{\rho}_t\right]
  + \frac{\kappa}{2S}\Big( \hat{S}_A^{-} \, \hat{\rho}_t \, \hat{S}_A^{+} + \hat{S}_B^{+} \, \hat{\rho}_t \, \hat{S}_B^{-} - \tfrac{1}{2} \{ \hat{S}_A^{+}\hat{S}_A^{-}+\hat{S}_B^{-}\hat{S}_B^{+},\hat{\rho}_t \} \Big)\,,
  \label{eq:lind}
\end{equation}
where $\kappa$ is the decay (excitation) rate for subsystem $A$ ($B$). This dynamics conserves the variables $({\hat{S}_j})^2 = ({\hat{S}_j^x})^2+({\hat{S}_j^y})^2+({\hat{S}_j^z})^2$, with $j=A,B$, and -- as mentioned above -- we choose $({\hat{S}_j})^2 = S(S+1) = \frac{N}{2}\left(\frac{N}{2}+1\right)$ for both $j=A,\,B$. We fix $\kappa = 1$ throughout. 

For any generic operator $\hat{\mathcal{O}}$, we can write the Heisenberg-like evolution of its expectation value as~\cite{PhysRevLett.121.035301}
\begin{align}
  \frac{\ud}{\ud t}\braket{\hat{\mathcal{O}}}_t &= i\braket{[\hat{H},\hat{\mathcal{O}]}}_t \nonumber\\
  &+ \frac{\kappa}{2S}\braket{\left\{
      \left[\hat{S}_A^{+},\hat{\mathcal{O}}\right]\hat{S}_A^{-}+
      \hat{S}_A^{+}\left[\hat{\mathcal{O}},\hat{S}_A^{-}\right] + 
      \left[\hat{S}_B^{-},\hat{\mathcal{O}}\right]\hat{S}_B^{+}+
      \hat{S}_B^{-}\left[\hat{\mathcal{O}},\hat{S}_B^{+}\right]\right\}}_t\,,
\end{align}
where we have defined $\braket{\ldots}_t = \Tr[(\ldots)\hat{\rho}_t]$. Considering $\hat{\mathcal{O}} = \hat{S}_j^\alpha$ with $\alpha=x,y,z$ we get the evolution of the expectations of the components of the subsystem magnetizations
\begin{equation}\label{heiso:eqn}
\begin{aligned}
  \frac{\ud}{\ud t}\braket{\hat{S}_A^x}_t &= \frac{\Gamma}{S}\braket{\hat{S}_A^z\hat{S}_B^{y}}+\frac{\kappa}{2S}\left(\{\hat{S}_A^z,\hat{S}_A^x\}-\hat{S}_A^x\right)\\
    \frac{\ud}{\ud t}\braket{\hat{S}_B^x}_t &= \frac{\Gamma}{S}\braket{\hat{S}_B^z\hat{S}_A^{y}}-\frac{\kappa}{2S}\left(\{\hat{S}_B^z,\hat{S}_B^x\}-\hat{S}_B^x\right)\\
  \frac{\ud}{\ud t}\braket{\hat{S}_A^y}_t &= -\Omega\braket{\hat{S}_A^z}_t-\frac{\Gamma}{S}\braket{\hat{S}_A^z\hat{S}_B^{x}}+\frac{\kappa}{2S}\left(\{\hat{S}_A^z,\hat{S}_A^y\}-\hat{S}_A^y\right)\\
  \frac{\ud}{\ud t}\braket{\hat{S}_B^y}_t &= -\Omega\braket{\hat{S}_B^z}_t-\frac{\Gamma}{S}\braket{\hat{S}_B^z\hat{S}_A^{x}}-\frac{\kappa}{2S}\left(\{\hat{S}_B^z,\hat{S}_B^y\}-\hat{S}_B^y\right)\\
  \frac{\ud}{\ud t}\braket{\hat{S}_A^z}_t &= \Omega\braket{\hat{S}_A^y}_t+\frac{\Gamma}{2S}\left(\hat{S}_A^y\hat{S}_B^x-\hat{S}_A^x\hat{S}_B^y\right)-\frac{\kappa}{S}\left[(\hat{S}_A^x)^2+(\hat{S}_A^y)^2\right]\\
  \frac{\ud}{\ud t}\braket{\hat{S}_B^z}_t &= \Omega\braket{\hat{S}_B^y}_t+\frac{\Gamma}{2S}\left(\hat{S}_B^y\hat{S}_A^x-\hat{S}_B^x\hat{S}_A^y\right)+\frac{\kappa}{S}\left[(\hat{S}_B^x)^2+(\hat{S}_B^y)^2\right]\,.
\end{aligned}
\end{equation}
If we consider the reduced variables $\hat{m}_{A/B}^\alpha = \hat{S}_{A/B}^\alpha / S$ we see that they commute in the limit $S\to\infty$. {The key point is that the $\hat{m}_j^\alpha$ variables obey the commutation rule $[\hat{m}_j^\alpha,\hat{m}_l^\beta]=i \delta_{jl}\epsilon^{\alpha\beta\gamma}\hat{m}_j^\gamma / S$, so that in the limit $S\to\infty$ they commute and the dynamics is classical.} In this limit, therefore, one can neglect quantum correlations for these variables. So, rewriting Eq.~\eqref{heiso:eqn} for the reduced variables, taking the limit $S\to\infty$ and neglecting quantum correlations, one obtains
\begin{equation}\label{dimerm:eqn}
\begin{aligned}
\frac{\ud m_{A}^{x}}{\ud t} &= \kappa \, m_{A}^{x} m_{A}^{z} + \Gamma \, m_{A}^{z} m_{B}^{y} \\
\frac{\ud m_{B}^{x}}{\ud t} &= -\kappa \, m_{B}^{x} m_{B}^{z} + \Gamma \, m_{B}^{z} m_{A}^{y} \\
\frac{\ud m_{A}^{y}}{\ud t} &= -\Omega \, m_{A}^{z} + \kappa \, m_{A}^{y} m_{A}^{z} - \Gamma \, m_{A}^{z} m_{B}^{x} \\
\frac{\ud m_{B}^{y}}{\ud t} &= -\Omega \, m_{B}^{z} - \kappa \, m_{B}^{y} m_{B}^{z} - \Gamma \, m_{B}^{z} m_{A}^{x} \\
\frac{\ud m_{A}^{z}}{\ud t} &= \Omega \, m_{A}^{y} - \kappa \left( (m_{A}^{x})^2 + (m_{A}^{y})^2 \right) + \Gamma \left( m_{A}^{y} m_{B}^{x} - m_{A}^{x} m_{B}^{y} \right) \\
\frac{\ud m_{B}^{z}}{\ud t} &= \Omega \, m_{B}^{y} + \kappa \left( (m_{B}^{x})^2 + (m_{B}^{y})^2 \right) + \Gamma \left( m_{B}^{y} m_{A}^{x} - m_{B}^{x} m_{A}^{y} \right).
\end{aligned}
\end{equation}

\subsection{Quantum-trajectory approach for the finite-size case}\label{trajectory:sec}

Let us consider Eq.~\eqref{eq:lind} where, as stated above, we restrict to the tensor product subspace such that $({\hat{S}_A})^2 = ({\hat{S}_B})^2 = S(S+1)$. Since the dynamics conserves both $({\hat{S}_A})^2 $ and $({\hat{S}_B})^2$, we can fix these values throughout the evolution. Therefore we can notice that 
\begin{equation}
\begin{aligned}
  \hat{S}_A^+\hat{S}_A^- &= - ({\hat{S}_A^z})^2+({\hat{S}_A})^2+\hat{S}_A^z = S\left(S+1\right) + \hat{S}_A^z - ({\hat{S}_A^z})^2\\
  \hat{S}_B^-\hat{S}_B^+ & = S\left(S+1\right) - \hat{S}_B^z - ({\hat{S}_B^z})^2\,.
\end{aligned}
\end{equation}
We implement this dynamics using the quantum-jump unraveling of the Lindblad equation Eq.~\eqref{eq:lind}. One can show that the Lindblad evolution is equivalent to the average over infinite realizations of a pure-state dynamics with a stochastic non-Hermitian Schr\"odinger equation. For each Lindblad equations there are many possible stochastic-dynamics approaches leading to it and are called unravelings (see Refs.~\cite{Plenio,Daley2014,fazio2025manybodyopenquantumsystems} for a review). Here we choose the so-called quantum-jump unraveling that takes the following form. We discretize the time dividing it in intervals $\delta t$ and in each time interval we proceed as follows.
\begin{itemize}
  \item With probability
  \begin{equation}
    p_1 = \frac{\kappa}{2S}\delta t\braket{\psi_t|\hat{S}_A^+\hat{S}_A^-|\psi_t} = \frac{\kappa}{2S}\delta t\left[S\left(S+1\right) + \braket{\psi_t|[\hat{S}_A^z-(\hat{S}_A^z)^2]|\psi_t}\right]
  \end{equation}
  we perform the transformation
  \begin{equation}
    \ket{\psi_t} \to \ket{\psi_{t+\delta t}} = \frac{\hat{S}_A^-\ket{\psi_t}}{||\hat{S}_A^-\ket{\psi_t}||}\,.
  \end{equation}
  \item With probability
  \begin{equation}
    p_2 = \frac{\kappa}{2S}\delta t\braket{\psi_t|\hat{S}_B^-\hat{S}_B^+|\psi_t} = \frac{\kappa}{2S}\delta t\left[S\left(S+1\right) - \braket{\psi_t|[\hat{S}_B^z+(\hat{S}_B^z)^2]|\psi_t}\right]
  \end{equation}
  we perform the transformation
  \begin{equation}
    \ket{\psi_t} \to \ket{\psi_{t+\delta t}} = \frac{\hat{S}_B^+\ket{\psi_t}}{||\hat{S}_B^+\ket{\psi_t}||}\,.
  \end{equation}
  \item With probability $p = 1 -p_1 -p_2$ we evolve along $\delta t$ with the non-Hermitian Hamiltonian
  \begin{equation}
    \hat{H}_{\rm nH} = \Omega ( \hat{S}_A^x + \hat{S}_B^x ) + \frac{\Gamma}{2S}\left(\hat{S}_A^+\hat{S}_B^-+\hat{S}_A^-\hat{S}_B^+\right) -i\frac{\kappa}{4S}\left(\hat{S}_A^+\hat{S}_A^-+\hat{S}_B^-\hat{S}_B^+\right)\,,
  \end{equation}
  or equivalently with
  \begin{equation}
    \hat{H}_{\rm nH}' = \Omega ( \hat{S}_A^x + \hat{S}_B^x ) + \frac{\Gamma}{2S}\left(\hat{S}_A^+\hat{S}_B^-+\hat{S}_A^-\hat{S}_B^+\right) -i\frac{\kappa}{4S}\left(\hat{S}_A^z-\hat{S}_B^z-(\hat{S}_A^z)^2-(\hat{S}_B^z)^2\right)\,.
  \end{equation}
  (The two Hamiltonians are equal up to a constant.) Writing $\hat{H}_{\rm nH}' = \hat{H} -i \hat{K}$, we evolve by trotterization
  \begin{equation}
    \ket{\psi_t} \to \ket{\psi_{t+\delta t}} = \frac{\nep^{-i\hat{H}\delta t}\nep^{\hat{K}\delta t}\ket{\psi_t}}{||\nep^{-i\hat{H}\delta t}\nep^{\hat{K}\delta t}\ket{\psi_t}||}\,.
  \end{equation}
\end{itemize}
We initialize the system with both collective spins pointing upwards, $S_A^z=S_B^z= S$ and $S_A=S_B=S$, and fix everywhere $\kappa=1$. 

\section{Mean-field dynamics}\label{dim:sec}

\subsection{Methods}

\subsubsection{Equations.}

Considering the infinite spin-magnitude limit $S\to\infty$, one gets a mean-field dynamics, described by Eq.~\eqref{dimerm:eqn}. Exploiting the conservation of the quantities $||{\bf m}_j||^2\equiv (m_j^x)^2+(m_j^y)^2+(m_j^z)^2$, we restrict to the case  $||{\bf m}_A||^2=||{\bf m}_B||^2=1$. 

\subsubsection{Initialization.}\label{ini:sec}

Initial conditions are generated by taking $(m_j^x, m_j^y, m_j^z) = (e,\, d,\, \sqrt{1 - e^2 - d^2})$ with ($e$, $d$) uniformly sampled within a disk of radius $a$. The sampling satisfies the constraint
\begin{equation}\label{constraint:eqn}
  e^2 + d^2 \leq a^2.
\end{equation}
In polar coordinates we can write for all $j$ $(m^x, m^y, m^z) = (\sin\theta\cos\varphi, \sin\theta\sin\varphi, \cos\theta)$ and $0<\sin\theta<a$, that is to say $0<\theta<\arcsin a$. The variables $e$ and $d$ are jointly uniformly distributed over a disk of radius $a$, meaning their joint probability density function is
\begin{equation}
p(e, d) = \frac{1}{\pi a^2}\,.
\end{equation}
Therefore we average over a cap surrounding the north pole of the sphere. Notice that $a=1$ corresponds to averaging over the whole northern hemisphere.

\subsubsection{Largest Lyapunov exponent.}\label{lyapel:sec}

To evaluate the chaoticity of the dynamics we evaluate the largest Lyapunov exponent $\Lambda_L$~\cite{lyap} (LLE), defined as the limit of the logarithmic rate of separation of two initially close trajectories
\begin{equation}
  \Lambda_L = \lim_{d_0\to 0}\lim_{t \to \infty} \frac{1}{t} \ln \left[ \frac{d(t)}{d_0} \right]\,,
\end{equation}
where $d_0$ is the initial distance between the two trajectories, and $d(t)$ is the distance at time $t$. To compute the Lyapunov exponent numerically, we use the method described in Ref.~\cite{PhysRevA.14.2338}. We evolve two initially close trajectories according to Eq.~\eqref{dimerm:eqn}. We discretize the time with steps $\delta t$ and at each time step $n$, we calculate the distance $d_n$ between the two trajectories, and evaluate the logarithm of the ratio of $d_n$ and $d_0$. Then we move one of the phase points along the line joining the two points in order to restore the initial distance, and repeat the process. The set of obtained logarithms is then averaged over time to obtain the Lyapunov exponent
\begin{equation}
  \Lambda_L = \frac{\delta t}{T_{\rm max}}\sum_{n=1}^{T_{\rm max}/\delta t} \ln\left(\frac{d_n}{d_0}\right)\,.
\end{equation}
Specifically, we perform the averaging over a time interval of length $T_{\text{max}}$, call the number of steps used as $K\equiv T_{\rm max}/\delta t$, and average $\Lambda_L$ over a set of randomly chosen initial conditions, as specified in Sec.~\ref{ini:sec}. We mark the Lyapunov exponent averaged over random initial conditions as $\braket{\Lambda_L}$.

\subsubsection{Pearson coefficient.}\label{pear:sec}

To study the correlations between the subsystems, we calculate the Pearson coefficient $C_P$ (see for instance Refs.~\cite{Cabot_2019,Galve_2017}). To compute this coefficient for the time evolution of subsystems $A$ and $B$, we use the following approach. We take a set of random initial conditions as specified in Sec.~\ref{ini:sec}. At each time $t$ we evaluate the Pearson correlation coefficient between {\em different subsystems} as
\begin{equation}\label{cicot:eqn}
  C_P(t) = \frac{\braket{m_A^z(t)m_B^z(t)}-\braket{m_A^z(t)}\braket{m_B^z(t)}}{\sqrt{\braket{(m_A^z)^2(t)}-\braket{m_A^z(t)}^2}\sqrt{\braket{(m_B^z)^2(t)}-\braket{m_B^z(t)}^2}}\,,
\end{equation}
where $\braket{\cdots}$ marks the average over random initial conditions. We find that this quantity attains a steady-state value after an initial transient behavior. We average over time in the steady-state regime and obtain $\overline{C_P}$, where $\overline{(\cdots)}$ marks the time average.

\subsection{Results}

\begin{figure}[t]
    \centering
    \includegraphics[width=\textwidth]{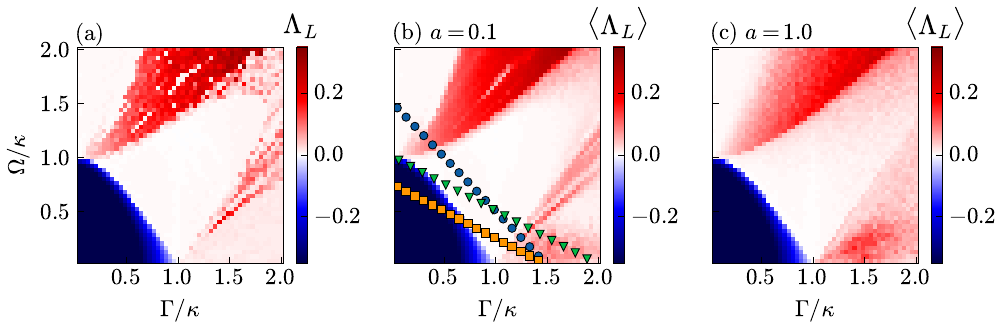}
    \caption{Largest Lyapunov exponent $\Lambda_L$ versus $\Gamma$ and $\Omega$ for initial conditions $m_j^z=1$ and $m_j^{x,y}=0$ for all $j$ (panel a), averaged over random initial conditions as given in Sec.~\ref{ini:sec} with $a=0.1$ (panel b), and with $a=1$ (panel c). The averages are over $N_{\rm r}=50$ initial conditions. See the main text and Fig.~\ref{fig:correlations_2} for a discussion about the symbols shown in panel (b).}
	\label{fig:all_imageslyap_chain}
\end{figure}

\begin{figure}[b]
	\centering
    \includegraphics[width=\textwidth]{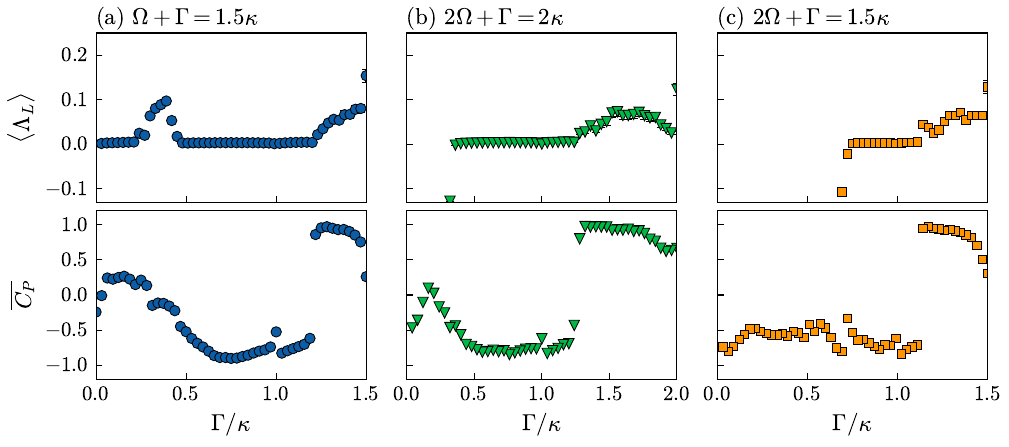}
	\caption{Largest Lyapunov exponent $\braket{\Lambda_L}$ (top row) and Pearson correlation coefficient $\overline{C_P}$ (bottom row) versus $\Gamma$ along the line $\Gamma + \Omega = 1.5$ (left column), $2\Omega + \Gamma = 2\kappa$ (center column), and $2\Omega + \Gamma = 1.5\kappa$ (right column). These values are calculated as averages over $N_{\rm r} = 100$ initial conditions generated on a spherical cap with parameter $a = 0.1$.}
	\label{fig:correlations_2}
\end{figure}

We begin by plotting the LLE defined in Sec.~\ref{lyapel:sec}, shown in Fig.~\ref{fig:all_imageslyap_chain}. In panel (a) we choose a specific initial condition (all spins pointing along the north pole), while in panels (b) and (c) we average over initial conditions as given in Sec.~\ref{ini:sec}, choosing respectively $a=0.1$ and $a=1$. All the regimes described in Ref.~\cite{solanki2024chaostimedissipativecontinuous} for Fig.~\ref{fig:all_imageslyap_chain}(a) are clearly visible also when we average over many initial conditions, though they appear smeared out. In particular, at the bottom left there is a regime of negative LLE (absence of chaos) corresponding to relaxation of the two magnetizations to a fixed point (trivial phase). Then there is a central lobe where the LLE is vanishing (again, absence of chaos) that corresponds to a continuous time crystal phase (marked as CTC3 in Ref.~\cite{solanki2024chaostimedissipativecontinuous}). Another similar lobe appears in the upper left region, where there is another continuous time crystal phase (CTC1) which encompasses the usual dissipative time crystal (occurring when the coupling $\Gamma$ vanishes).

\begin{figure}[t]
    \centering
    \includegraphics[width=\textwidth]{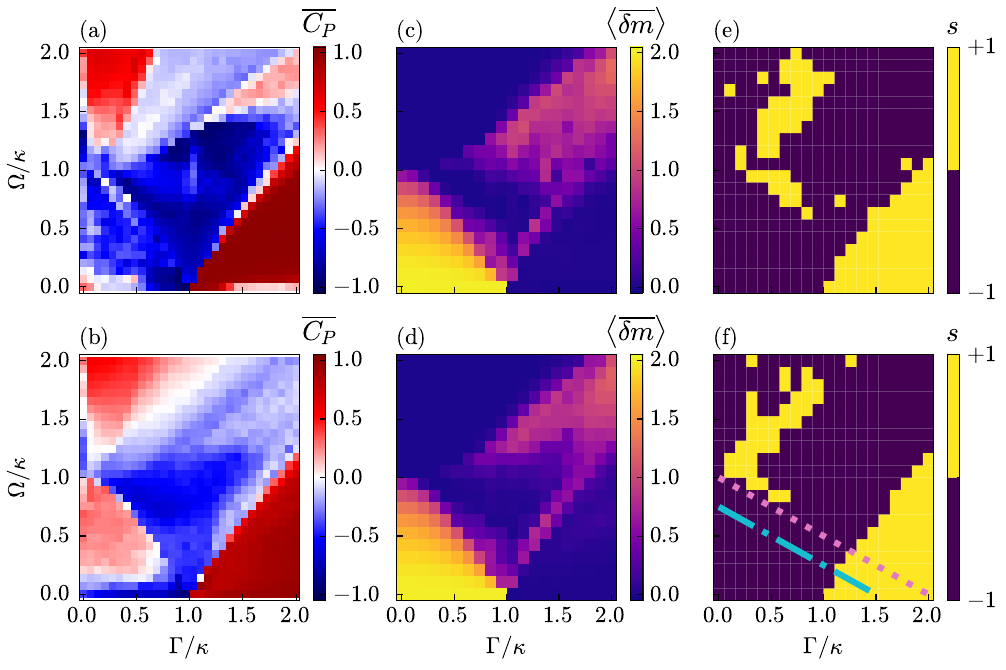}
    \caption{Heat maps of various quantities versus $\Gamma$ and $\Omega$. (a,b) Pearson correlation coefficient. (c,d) Difference between the magnetizations averaged over time and initial conditions, $\langle \overline{\delta m} \rangle = \braket{\overline{m_B^z}}-\braket{\overline{m_A^z}}$. (e,f) Relative sign $s = \sign(\braket{\overline{m_z^A}}\braket{\overline{m_z^B}})$ of the two magnetizations. The calculations are performed by averaging over $N_{\rm r}=100$ random initial conditions generated on a spherical cap with parameter $a = 0.1$ (first row), and $a = 1$ (second row). The lines in panel (f) are discussed in Sec.~\ref{qua:sec}.}
    \label{fig:pearson_heat_map}
\end{figure}

In Fig.~\ref{fig:all_imageslyap_chain}(b), we consider three lines cutting the parameter space in Fig.~\ref{fig:all_imageslyap_chain}(b) from the upper left to the bottom right part of the panel; These lines are $\Omega + \Gamma = 1.5\kappa$, $2\Omega + \Gamma = 2\kappa$, and $2\Omega + \Gamma = 1.5\kappa$ and are marked in Fig.~\ref{fig:all_imageslyap_chain}(b) with different symbols (blue circles, green triangles, and orange squares, respectively). Along these lines, we plot, using the same symbols, the LLE (Fig.~\ref{fig:correlations_2}, top row) and the corresponding average Pearson coefficient described in Sec.~\ref{pear:sec} (Fig.~\ref{fig:correlations_2}, bottom row) versus $\Gamma$. Both the LLE and the Pearson coefficient are evaluated over a set of random initial conditions with $a=0.1$. First of all, we notice in panels~\ref{fig:correlations_2}(c,d) that there is a regime of negative Lyapunov exponent corresponding to the trivial phase. Beyond that, we see that there is a threshold value [$\Gamma \simeq 1.2$ in panel~\ref{fig:correlations_2}(a), $\Gamma\simeq 1.25$ in panel~\ref{fig:correlations_2}(b), $\Gamma\simeq 1.15$ in panel~\ref{fig:correlations_2}(c)] where the LLE moves from zero to positive, coinciding with a sudden increase of the Pearson coefficient. At this threshold, the Pearson coefficient shifts from strongly negative to strongly positive values. Below this threshold the system is in the CTC3 phase: Here, the Lyapunov exponent is vanishing, and the negative Pearson coefficient indicates that the two subsystem oscillate out of phase with respect to each other. 
Above the threshold the system becomes chaotic -- as marked by the Lyapunov exponent becoming positive -- but the two oscillators behave in a strictly correlated way, as witnessed by the positive value of the Pearson coefficient. Therefore, in this regime, the two subsystems display full chaotic synchronization. 

The full chaotic synchronization regime is better highlighted in Fig.~\ref{fig:pearson_heat_map}(a,b), showing heat maps of the Pearson coefficient across the parameter space. In Fig.~\ref{fig:pearson_heat_map}(a) we show the heat map of the Pearson coefficient averaged over a cap with $a=0.1$ and we can see a close correspondence of this plot with the heat map of the LLE evaluated with the same initial conditions [see Fig.~\ref{fig:all_imageslyap_chain}(b)].

\begin{figure}[t]
	\centering
    \includegraphics[width = \textwidth]{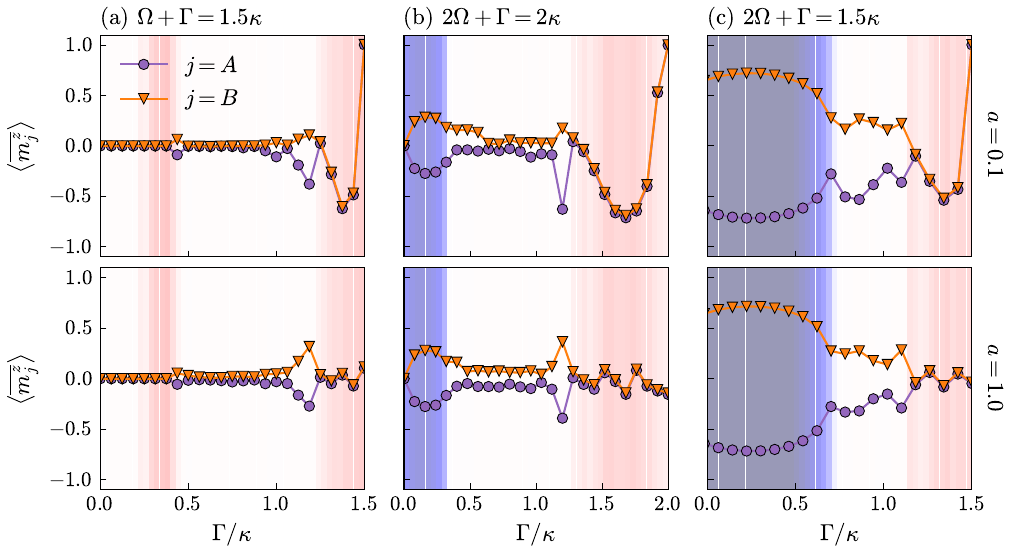}
	\caption{Time and initial-condition average values of $m_j^z(t)$, $\braket{\overline{m_j^z}}$ versus $\Gamma$, along the parameter-space line $\Omega + \Gamma = 1.5\kappa$ (a, left column), $2\Omega + \Gamma = 2\kappa$ (b, center column), and $2\Omega + \Gamma = 1.5\kappa$ (c, right column). We average over a set of $N_{\rm r} = 100$ initial conditions generated either for $a = 0.1$ (top row) or $a=1$ (bottom row). The legend is shared. The background shadowing marks the Lyapunov exponent with the same color code as Fig.~\ref{fig:all_imageslyap_chain}.} 
	\label{fig:avg_mz_2}
\end{figure}

The behaviors of the Lyapunov exponent and Pearson coefficient closely relate to the behavior of the magnetization averaged over time and initial conditions. We define this average as $\braket{\overline{m_j^z}}$ ($j=A,B$) and  in Fig.~\ref{fig:avg_mz_2} we plot it versus $\Gamma$  along the three parameter-space lines considered above. We identify many regimes. In panel~\ref{fig:avg_mz_2}(a) we see {\em (i)} a small-$\Gamma$ regime, where the average magnetization is vanishing for both subsystems; {\em (ii)} an intermediate-$\Gamma$ regime, where the average magnetizations of the two subsystems have opposite sign (staggered magnetization regime); {\em (iii)} a large-$\Gamma$ regime, where the average magnetizations of the two subsystems are nonvanishing and nearly equal (uniform magnetization regime). We find that regime {\em (i)} corresponds to the CTC1 phase, regime {\em (ii)} to the CTC3 regime and regime {\em (iii)} to the chaotic synchronized phase, as highlighted by the shadow scale in the background marking the value of the Lyapunov exponent. (Note that CTC1 and CTC3 are separated by a small chaotic region.) In particular, the crossover between the regime {\em (ii)} of staggered magnetization and regime {\em (iii)} of uniform magnetization coincides with the onset of full chaotic synchronization, as we can see by comparing the top row of Fig.~\ref{fig:avg_mz_2} (where initial conditions are with $a=0.1$) with Fig.~\ref{fig:correlations_2}. We can see regimes {\em (ii)} and {\em (iii)} also in panels~\ref{fig:avg_mz_2}(b,c), and also here the crossover between them corresponds to the onset of chaotic synchronization, as further highlighted by the shadowing that marks the value of the Lyapunov exponent. Here, regime {\em (i)} is absent, because there is no CTC1 phase along these lines in the parameter space, but there is the trivial phase instead. Here there is a staggered magnetization and the crossover to CTC3 is marked by a discontinuity in the derivative, closely mirroring the behavior of the Lyapunov exponent.

In order to better highlight these regimes, we plot in Fig.~\ref{fig:pearson_heat_map}(c,d) the heat maps for the difference between the magnetizations averaged over time and initial conditions, $\delta m \equiv\braket{\overline{m_B^z}}-\braket{\overline{m_A^z}}$ [in panel (c) we take $a=0.1$, in panel (d) $a=1$], and compare them with the  heat maps of the Pearson coefficient, with the corresponding value of $a$ [Fig.~\ref{fig:pearson_heat_map}(a,b)]. We see two regions where $\langle\overline{\delta m}\rangle \approx 0$, on the top left and the bottom right, both corresponding to positive values of the Pearson coefficient. The top-left region corresponds to regime {\em (i)} with vanishing magnetization, while the bottom-right region corresponds to regime {\em (iii)}, characterized by uniform magnetization. We see also a central lobe where the difference $\delta m$ is slightly positive and the Pearson coefficient is negative. This region corresponds to the staggered magnetization regime {\em (ii)}. We also plot the heat map of the relative sign of the two averaged magnetizations, i.\,e., $s = \sign(\braket{\overline{m_A^z}}\braket{\overline{m_B^z}})$ [Fig.~\ref{fig:pearson_heat_map}(e,f)], and see a clear positive region at the bottom right. This is the region where regime {\em (iii)} occurs and -- as we can see comparing with Fig.~\ref{fig:all_imageslyap_chain}(b,c) and Fig.~\ref{fig:pearson_heat_map}(a,b) -- it corresponds to a positive Lyapunov exponent and to a positive Pearson coefficient that suddenly increases at the boundary of this region, as we have described above. Thus, this is a region of chaotic synchronization, that is therefore closely related to the presence of a uniform magnetization.

\section{Quantum finite-size dynamics}\label{qua:sec}

 \begin{figure}[t]
 \centering
 \includegraphics[width=\textwidth]{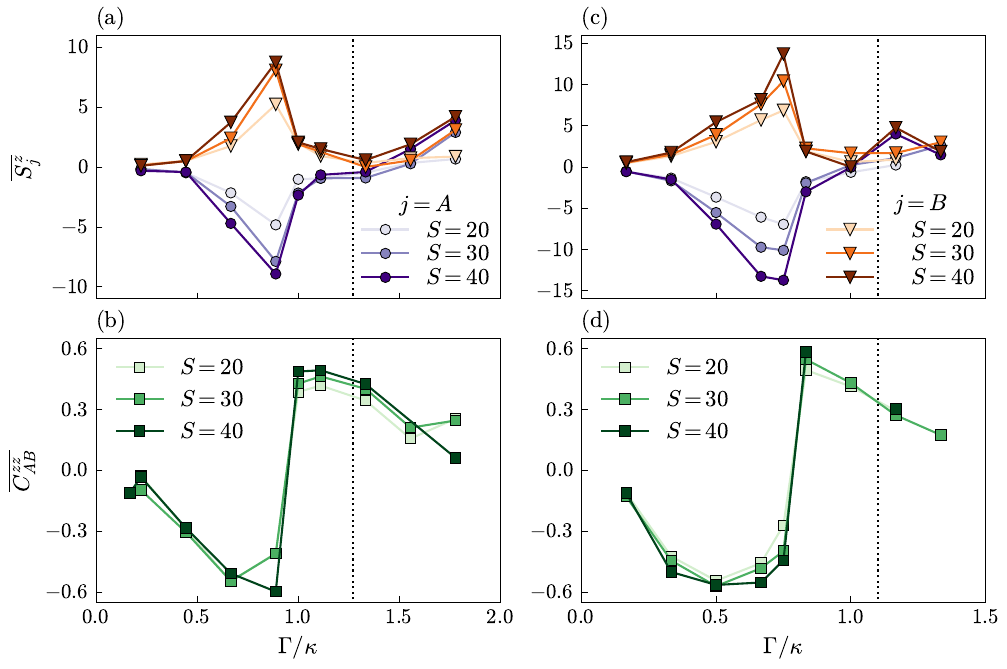}
  \caption{(a,c) $\overline{S_A^z}$ and $\overline{S_B^z}$ versus $\Omega$ for different $N$. (b,d) Time-averaged connected correlator $\overline{C_{AB}^{zz}}$. Average over $N_{\rm r } \geq 48$ trajectories. On the left column we consider the curves for the line $2\Omega + \Gamma = 2\kappa$ and on the right one $2\Omega + \Gamma = 1.5\kappa$. The vertical line marks the classical staggered-uniform crossover as found in Fig.~\ref{fig:pearson_heat_map}(f).}\label{medSz:fig}
 \end{figure}

In this section, we consider finite-size dynamics along quantum trajectories as explained in Sec.~\ref{trajectory:sec}. We restrict to the Hilbert subspace where the magnitudes of both spins are $S$, and always initialize the system in the state $\ket{S,S}$, with both spins pointing to the north pole. If $N_{\rm r}$ is the number of trajectories, the expectation values of the observables on the Lindblad density matrix are given by an average over trajectories
\begin{equation}
  \braket{\mathcal{O}}_t=\Tr[\hat{\mathcal{O}}\hat{\rho}_t] = \lim_{N_{\rm r}\to\infty}\frac{1}{N_\text{r}}\sum_{r=1}^{N_{\rm r}}\braket{\psi_t^r|\hat{\mathcal{O}}|\psi_t^r}\,,
\end{equation}
where $\ket{\psi_t^r}$ is the state along the $r$-the trajectory realization at time $t$. We plot the time averages $\overline{S_{A/B}^z}$ versus $\Gamma$ in Fig.~\ref{medSz:fig}. We plot them along the line $2\Omega + \Gamma= 2\kappa$ in Fig.~\ref{medSz:fig}(a) and along the line  $2\Omega + \Gamma= 1.5\kappa$ in Fig.~\ref{medSz:fig}(c). Also in this quantum regime, a staggered regime exists for small values of $\Gamma$, while above a threshold [$\Gamma\simeq 1.0$ in panel (a) and $\Gamma\simeq 0.8$ in panel (c)] the system moves to a regime characterized by more uniform magnetization, similarly to the classical case. The two regimes are visible also in the connected correlator that abruptly changes sign at the crossover point [Fig.~\ref{medSz:fig}(b,d)]:
\begin{equation}
  \overline{C_{AB}^{zz}} = \overline{\braket{S_A^zS_B^z}}-\overline{\braket{S_A^z}\braket{S_B^z}}.
\end{equation}
We see here that there is no correspondence between the quantum and the classical behavior, highlighted by the vertical line that marks the value of the staggered-uniform crossover along the considered parameter-space line, as read in Fig.~\ref{fig:pearson_heat_map}(f). 

\begin{figure}[t]
  \centering
  \includegraphics[width=\textwidth]{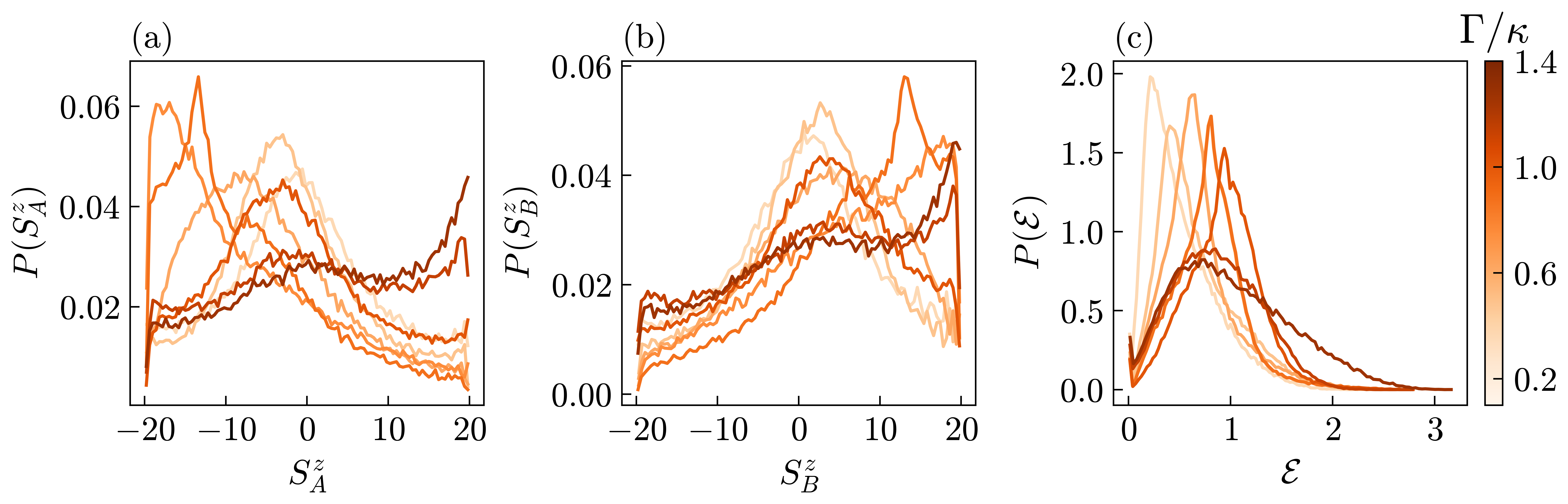}
  \caption{(a) Histogram of $(S_A^z)^r(t)$ as it changes along time $t$ and across trajectories $r$. (b) Histogram of $(S_B^z)^r(t)$ as it changes along time $t$ and across trajectories $r$. (c) Histogram of the entanglement entropy $\mathcal{E}^r(t)$ as it changes along time $t$ and across trajectories $r$. We fix the constraint $2\Omega + \Gamma = 1.5\kappa$, $\delta t = 10^{-4}$ and $S= 20$. }\label{diSt:fig}
\end{figure}

\begin{figure}[t]
    \centering
    \includegraphics[width=\textwidth]{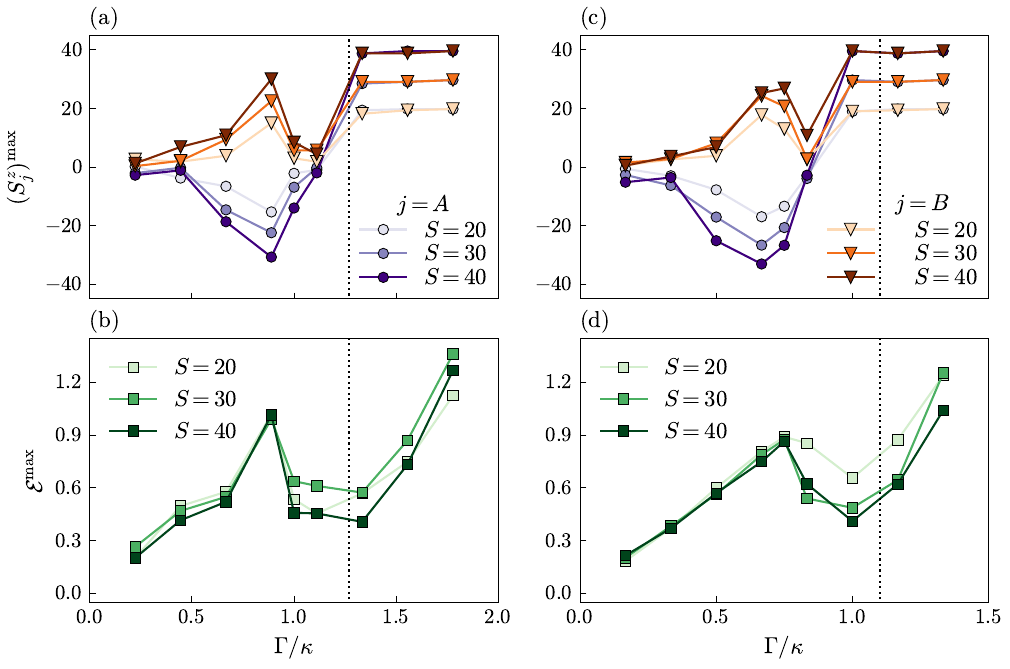}
  \caption{(a,c) Maximum points of the histograms of $S_A^z$ and $S_B^z$ along and across trajectories versus $\Gamma$. (b,d) Maximum point of the histogram of the entanglement entropy versus $\Gamma$. We fix the constraint $2\Omega + \Gamma = 2\kappa$ in the left column and $2\Omega + \Gamma = 1.5\kappa$ in the right column. We take also $\delta t = 10^{-4}$.}\label{St:fig}
\end{figure}

Further insight is gained by analyzing histograms of these quantities over time $t$ and over the ensemble of different trajectory realizations $r$ (as done in Refs.~\cite{gda_EPJB,Passarelli_2025}). We plot the histograms for $(S_A^z)^r(t) = \braket{\psi_t^r|\hat{S}_A^z|\psi_t^r}$ and $(S_B^z)^r(t) = \braket{\psi_t^r|\hat{S}_B^z|\psi_t^r}$ in Fig.~\ref{diSt:fig}(a,b). They are obtained as an histogram over a set of values obtained by changing $r$ and $t$. Along each trajectory, we also consider the entanglement entropy between the two subsystems, defined as
\begin{equation}
  \mathcal{E}^r(t) = -\Tr\left[\hat{\rho}_A^r(t)\ln\hat{\rho}_A^r(t)\right]\quad\text{with}\quad \hat{\rho}_A^r(t)=\Tr_B\left[\ket{\psi_t^r}\bra{\psi_t^r}\right]\,,
\end{equation}
and plot its histograms in Fig.~\ref{diSt:fig}(c). All histograms have a maximum, and we plot its position as a function of $\Gamma$ in Fig.~\ref{St:fig}. We can clearly observe the crossover from the staggered to the uniform magnetized regime in the behavior of $(S_A^z)^{\rm max}$ and $(S_B^z)^{\rm max}$ [Fig.~\ref{St:fig}(a,c)], also accompanied by a local minimum in the behavior of the maximum point of the entanglement entropy $\mathcal{E}^{\text{max}}$ [Fig.~\ref{St:fig}(b,d)].

We notice that the threshold for this crossover in the quantum case is in general not consistent with the mean field case, as we can see comparing Fig.~\ref{St:fig}(a,c) with Fig.~\ref{fig:pearson_heat_map}(f). We see this fact comparing the curves with the vertical line in the plots that marks the classical crossover from the staggered-magnetization to the uniform-magnetization regime, as read from~\ref{fig:pearson_heat_map}(f). While along the parameter-space line $2\Omega +\Gamma = 2\kappa$ there is a reasonable correspondence (left column of Fig.~\ref{St:fig}), this is not true along a different parameter-space line ($2\Omega + \Gamma = 1.5\kappa$ -- left column of Fig.~\ref{St:fig}. Nevertheless, the quantum behavior is qualitatively similar to the classical one, because in both Fig.~\ref{fig:avg_mz_2} and in Fig.~\ref{diSt:fig} one sees a similar staggered-uniform crossover.


Because the classical system’s crossover closely relates to full chaotic synchronization, we interpret these quantum observations as evidence of quantum synchronization. In analogy with the classical mean-field case, we can say that the quantum synchronization sets in when the magnetization suddenly becomes uniform. Moreover, as in the classical case, also in the quantum one the synchronization is chaotic, as one can argue considering the properties of the average density matrix
\begin{equation}\label{rhasi:eqn}
  \hat{\rho}_t = \frac{1}{N_{\rm r}}\sum_{r=1}^{N_{\rm r}} \ket{\psi_t^r}\bra{\psi_t^t}
\end{equation}
for a time long enough that the steady state has been reached. This is the so-called nonequilibrium steady state (NESS) density matrix, $\hat{\rho}_{\rm NESS}$. After obtaining $\hat{\rho}_{\rm NESS}$ as an average over realizations and {over a time window $[t_{\rm i},t_{\rm f}]$, choosing $t_{\rm i}$ such that} all finite-time transients have died out, we evaluate the average level spacing ratio $r_S$ of this matrix as described in Ref.~\cite{rufo2025quantumsemiclassicalsignaturesdissipative}. Denoting $\rho_{\alpha}$ the eigenvalues of $\hat{\rho}_{\rm NESS}$ taken in increasing order, the average level spacing ratio is defined as
\begin{equation}
	r_S = \frac{1}{2S-1}\sum_{\alpha=1}^{2S-1}\frac{\min(\rho_{\alpha+1}-\rho_{\alpha},\rho_{\alpha+2}-\rho_{\alpha+1})}{\max(\rho_{\alpha+1}-\rho_{\alpha},\rho_{\alpha+2}-\rho_{\alpha+1})}\,,
\end{equation}
{with the $\rho_\alpha$ taken in increasing order.} Evaluating $r_S$ (see Fig.~\ref{rhasi:fig}), we always find a value only slightly smaller than the random-matrix Gaussian unitary ensemble value $r_{\rm GUE} \simeq 0.5996$ resulting from a numerical fit in Ref.~\cite{Atas_2013}. {The reason behind the GUE statistics is that, due to the lack of time-inversion symmetry, there is in general no basis where it is fully real.}

{We emphasize that using the eigenvalues of $\hat{\rho}_{\rm NESS}$ -- as we do -- is equivalent to using the eigenvalues of $\hat{H}_{\rm eff}=-\ln\hat{\rho}_{\rm NESS}$ as done in~\cite{rufo2025quantumsemiclassicalsignaturesdissipative}. The point is that the difference between nearby eigenvalues of $\hat{\rho}_{\rm NESS}$ is very small $\sim 1/(2S+1)\ll 1$ and one can perform a linear expansion
 \begin{equation}
   |\rho_{\alpha+1} - \rho_\alpha| = |\nep^{-H_{\alpha+1}}-\nep^{-H_\alpha}| \simeq \nep^{-H_\alpha}|H_{\alpha +1} - H_\alpha|\,.
 \end{equation}
 When evaluating the level spacing ratio the factors $\nep^{-H_\alpha}$ simplify. It is the same principle why no unfolding is needed when evaluating the level spacing ratio. The approximation is very good already for $S=10$, and the two results agree within the errorbar, as we can see in Fig.~\ref{rhasi:fig}.}
 
 We can conclude that the steady-state Lindblad density matrix behaves as a random matrix, a behavior that corresponds to quantum chaos~\cite{rufo2025quantumsemiclassicalsignaturesdissipative}. Beyond that, it displays a sudden crossover from staggered to uniform magnetization strictly reminiscent of the synchronization phenomenology observed in the mean field [see Fig.~\ref{St:fig}(a,c)]. Therefore, also in the quantum finite-$S$ case, we can identify this phenomenon as quantum chaotic synchronization.

The fact that the quantum and the classical crossover do not coincide is related with the noncommutativity of the infinite-$S$ limit and the infinite time-limit. Indeed, the infinite-$S$ limit shows persistent (periodic or aperiodic) oscillations, while the finite-$S$ size case relaxes for $t\to\infty$ to a NESS density matrix for any arbitrary large but finite $S$~\cite{solanki2024chaostimedissipativecontinuous}. 
{{Indeed, our system has a well-defined semiclassical limit obtained by taking $S\to\infty$. In this limit, the mean-field equations exhibit a rich phenomenology at long times: convergence to stationary values (stationary phase), time-periodic oscillations (time-crystal phase), or irregular dynamics (chaotic phase).}

{{Non trivial long-time dynamics can emerge only when the thermodynamic limit is taken \emph{first}. In this scenario ($\lim_{t\to\infty} \lim_{S\to\infty}$), the mean-field system is described by the classical nonlinear equations Eq.~\eqref{dimerm:eqn} that can sustain  persistent oscillations or chaos. By contrast, for any finite $S$, quantum fluctuations and the finite-dimensional Hilbert space cause all dynamics to eventually converge to a steady state, after a transient behavior. Taking the long-time limit first ($\lim_{S\to\infty} \lim_{t\to\infty}$) thus washes out the interesting dynamical features before one moves to the thermodynamic limit.}

{{This subtle issue regarding the non-commutativity of limits has been discussed in recent literature on dissipative quantum systems. One example is Ref.~\cite{Delmonte_2025}, which analyzes dissipative quantum models that like ours are described by classical equations in the thermodynamic limit, and explicitly demonstrates how exchanging the order of these limits affects the phase diagram. In another work, quantum chaos in the driven-dissipative Bose-Hubbard dimer does not match the long-time regular behavior found in the classical limit~\cite{PhysRevResearch.7.013276}. Also this lack of quantum-classical correspondence can be interpreted as a noncommutativity of the thermodynamic and the long-time limits. Therefore, quantitative agreement remains inherently limited by the fundamental non-commutativity of the two limits.} 

Here we are studying the properties of the finite-$S$ NESS density matrix, which -- consistently with the discussion above -- are in general different from the ones of the long-time behavior in the $S\to\infty$ mean-field regime~\cite{Delmonte_2025,PhysRevResearch.7.013276}. So it is remarkable that the crossover is present in both scenarios, despite the non-commutativity of the limits and their different long-time behaviors. Additionally, entanglement plays a significant role in the quantum case but is classically absent. 

We also note that the noncommutative nature of these limits helps explain why the largest Lyapunov exponent, shown in Fig.~\ref{fig:correlations_2}, exhibits intervals of zero (regular dynamics) and positive (chaotic dynamics), whereas the NESS density matrix displays signatures of quantum chaos throughout the same parameter range (see Fig.~\ref{rhasi:fig}). This difference arises from the LLE assessing the long-time dynamics of the mean-field, infinite-$S$ limit, whereas the NESS characterizes the long-time limit of finite-size quantum dynamics -- the infinite-$S$ limit being taken first in the former case. {We emphasize that a similar behavior has been observed in a different model in~\cite{PhysRevResearch.7.013276}.}

\begin{figure}[t]
 \centering  
 \includegraphics[width=0.6\textwidth]{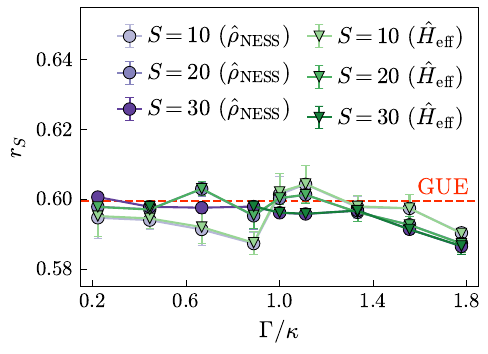}
	\caption{{The average level spacing ratio evaluated -- as explained in the text -- for the eigenvalues of $\hat{\rho}_{\rm NESS}$ and $\hat{H}_{\rm eff}$, and plotted versus $\Gamma$ along the line $2\Gamma+\Omega=2\kappa$, taking different values of $S$. The $\hat{\rho}_{\rm NESS}$ is evaluated using Eq.~\eqref{rhasi:eqn}, averaging over $N_{\rm r} = 48$ realizations and over a time window from $t_{\rm i} = 80$ to $t_{\rm f} = 100$, and then the average level spacing ratio $r_S$ is evaluated. $r_S$ is further averaged over $5\leq N_{\rm s} \leq 10$ different sets of disorder realizations, and the errorbar is evaluated as the root mean square deviation over sets divided by $\sqrt{N}_{\rm s}$.} }\label{rhasi:fig}
\end{figure}

\section{Conclusions}\label{con:sec}

In conclusion, we have investigated the dynamics of two coupled dissipative time crystals, identifying signatures of chaotic synchronization in both the classical and quantum regimes. In the mean-field infinite-$S$ limit, we found a region of parameter space characterized by a positive largest Lyapunov exponent and a Pearson correlation coefficient close to one, signaling the coexistence of chaos and strong synchronization~\cite{pecora_carro,PhysRevLett.71.65,strogatz:2000}. At the onset of this regime, the Pearson coefficient increases abruptly, in correspondence with a transition in the structure of the time-averaged magnetizations. Specifically, chaotic synchronization is associated with a uniform nonvanishing magnetization across the subsystems, whereas the staggered-magnetization regime corresponds to a  regular-dynamics time-crystal phases (CTC3 phase of Ref.~\cite{solanki2024chaostimedissipativecontinuous}), and the vanishing-magnetization one encompasses to a chaotic unsynchronized phase and another (CTC1) time crystal phase. Both the CTC1 and CTC3 phases display a nonvanishing Pearson coefficient (and then synchronization), but in those phases the Lyapunov exponent is vanishing and there is no chaos.  The presence of both chaos and nonvanishing Pearson coefficient appears only in the uniform-magnetization regime, and this establishes a direct correspondence between chaotic synchronization and collective order.

We then turned to finite-$S$ systems, where quantum fluctuations become relevant. Using a quantum-trajectory approach, we analyzed subsystem magnetizations along stochastic trajectories and their distributions over time and trajectory realizations. The maxima of these histograms exhibit a sharp crossover between staggered and uniform magnetization, closely resembling the mean-field case and indicating a quantum analogue of chaotic synchronization. Moreover, dissipative quantum chaos is revealed by Gaussian unitary ensemble statistics in the nonequilibrium steady-state density matrix~\cite{rufo2025quantumsemiclassicalsignaturesdissipative}. Beyond that, the entanglement entropy between subsystems displays a correlated feature (a local minimum) at the crossover, showing that entanglement contributes to the quantum synchronization process~\cite{Lee_2014,PhysRevLett.111.103605,Giorgi_2013}.

In summary, this model displays classical chaotic synchronization and an analogous quantum phenomenon, which we can call quantum chaotic synchronization. Interestingly, the classical crossover point between the staggered and the uniform magnetization -- where chaotic synchronization sets in -- does not in general coincide with the corresponding quantum one.
This is related to the fact that the limits $S\to\infty$ and $t\to\infty$ do not commute with each other. In the mean-field case, the $S\to\infty$ limit is taken first, resulting in persistent oscillations that can be periodic or aperiodic~\cite{solanki2024chaostimedissipativecontinuous}. In the quantum case, the $t\to\infty$ limit comes first and the system always relaxes to a NESS. Remarkably, the NESS and the mean-field dynamics share a qualitatively similar chaotic synchronization behavior, despite their fundamental differences and the presence of entanglement only in the quantum case. The noncommutativity of these limits furthermore explains the discrepancies in parameter regimes where classical and quantum chaos appear.

Our findings provide a consistent picture of chaotic synchronization between dissipative time crystals, in both the classical and the quantum regime, and support the identification of a phenomenon of quantum chaotic synchronization. Future research developments include the extension of this analysis to the case of a chain of many coupled subsystems (possibly with long-range dissipation~\cite{PhysRevB.106.224308}), the study of classical and quantum correlation in the case of classical and quantum coupled dissipative kicked rotors~\cite{PhysRevB.108.094305}, and the exploration of the relationship between chaotic synchronization and other kinds of quantum resources~\cite{d7tm-9hkp}.

\ack{We acknowledge P.\,Solanki for fruitful discussions. G.\,P. and A.\,R. acknowledge financial support from PNRR MUR Project No.~PE0000023-NQSTI. We acknowledge computational resources from the CINECA award under the ISCRA initiative and from MUR, PON “Ricerca e Innovazione 2014-2020,” under Grant No.~PIR01$\_$00011 - (I.Bi.S.Co.). This work was supported by PNRR MUR Project No.~PE0000023-NQSTI, by the European Union’s Horizon 2020 research and innovation program under Grant
Agreement No.~101017733, by MUR Project No.~CN$\_$00000013-ICSC (P.\,L.), and by the QuantERA II Programme STAQS project, which received funding from the European Union’s Horizon 2020 research and innovation program under Grant Agreement No.~101017733 (P.\,L.). E.\,P. acknowledges financial support from the Erasmus+ AURORA ALLIANCE fellowship program.
E.\,P. acknowledges support of the Czech Science Foundation (project 25-17472S). E.\,P. acknowledges project IGA PrF-2025-010. E.\,P. acknowledges use of the computational cluster of the Department of Optics, Palacký University.}

\bibliography{biblio}

@misc{solanki2024chaostimedissipativecontinuous,
      title={Chaos in Time: A Dissipative Continuous Quasi Time Crystals}, 
      author={Parvinder Solanki and Fabrizio Minganti},
      year={2024},
      eprint={2411.07297},
      archivePrefix={arXiv},
      primaryClass={quant-ph},
      url={https://arxiv.org/abs/2411.07297}, 
}

@article{PhysRevLett.111.103605,
  title = {Measures of Quantum Synchronization in Continuous Variable Systems},
  author = {Mari, A. and Farace, A. and Didier, N. and Giovannetti, V. and Fazio, R.},
  journal = {Phys. Rev. Lett.},
  volume = {111},
  issue = {10},
  pages = {103605},
  numpages = {5},
  year = {2013},
  month = {Sep},
  publisher = {American Physical Society},
  doi = {10.1103/PhysRevLett.111.103605},
  url = {https://link.aps.org/doi/10.1103/PhysRevLett.111.103605}
}

@article{Giorgi_2013,
   title={Spontaneous synchronization and quantum correlation dynamics of open spin systems},
   volume={88},
   ISSN={1094-1622},
   url={http://dx.doi.org/10.1103/PhysRevA.88.042115},
   DOI={10.1103/physreva.88.042115},
   number={4},
   journal={Physical Review A},
   publisher={American Physical Society (APS)},
   author={Giorgi, G. L. and Plastina, F. and Francica, G. and Zambrini, R.},
   year={2013},
   month=oct }

@article{PhysRevA.95.043807,
  title = {Quantum synchronization as a local signature of super- and subradiance},
  author = {Bellomo, B. and Giorgi, G. L. and Palma, G. M. and Zambrini, R.},
  journal = {Phys. Rev. A},
  volume = {95},
  issue = {4},
  pages = {043807},
  numpages = {11},
  year = {2017},
  month = {Apr},
  publisher = {American Physical Society},
  doi = {10.1103/PhysRevA.95.043807},
  url = {https://link.aps.org/doi/10.1103/PhysRevA.95.043807}
}

@Article{SciPostPhys.12.3.097,
	title={{Algebraic theory of quantum synchronization and limit cycles under dissipation}},
	author={Berislav Buča and Cameron Booker and Dieter Jaksch},
	journal={SciPost Phys.},
	volume={12},
	pages={097},
	year={2022},
	publisher={SciPost},
	doi={10.21468/SciPostPhys.12.3.097},
	url={https://scipost.org/10.21468/SciPostPhys.12.3.097},
}

@article{PhysRevLett.131.190402,
  title = {Macroscopic Quantum Synchronization Effects},
  author = {Nadolny, Tobias and Bruder, Christoph},
  journal = {Phys. Rev. Lett.},
  volume = {131},
  issue = {19},
  pages = {190402},
  numpages = {6},
  year = {2023},
  month = {Nov},
  publisher = {American Physical Society},
  doi = {10.1103/PhysRevLett.131.190402},
  url = {https://link.aps.org/doi/10.1103/PhysRevLett.131.190402}
}

@article{PhysRevA.109.033718,
  title = {Exploring quantum synchronization with a composite two-qubit oscillator},
  author = {Vaidya, Gaurav M. and Mamgain, Arvind and Hawaldar, Samarth and Hahn, Walter and Kaubruegger, Raphael and Suri, Baladitya and Shankar, Athreya},
  journal = {Phys. Rev. A},
  volume = {109},
  issue = {3},
  pages = {033718},
  numpages = {16},
  year = {2024},
  month = {Mar},
  publisher = {American Physical Society},
  doi = {10.1103/PhysRevA.109.033718},
  url = {https://link.aps.org/doi/10.1103/PhysRevA.109.033718}
}

@article{PhysRevLett.133.260403,
  title = {Exotic Synchronization in Continuous Time Crystals Outside the Symmetric Subspace},
  author = {Solanki, Parvinder and Krishna, Midhun and Hajdu\ifmmode \check{s}\else \v{s}\fi{}ek, Michal and Bruder, Christoph and Vinjanampathy, Sai},
  journal = {Phys. Rev. Lett.},
  volume = {133},
  issue = {26},
  pages = {260403},
  numpages = {6},
  year = {2024},
  month = {Dec},
  publisher = {American Physical Society},
  doi = {10.1103/PhysRevLett.133.260403},
  url = {https://link.aps.org/doi/10.1103/PhysRevLett.133.260403}
}

@article{Cabot_2019,
   title={Quantum Synchronization in Dimer Atomic Lattices},
   volume={123},
   ISSN={1079-7114},
   url={http://dx.doi.org/10.1103/PhysRevLett.123.023604},
   DOI={10.1103/physrevlett.123.023604},
   number={2},
   journal={Physical Review Letters},
   publisher={American Physical Society (APS)},
   author={Cabot, Albert and Giorgi, Gian Luca and Galve, Fernando and Zambrini, Roberta},
   year={2019},
   month=jul }

@article{Shen_2023,
   title={Quantum synchronization effects induced by strong nonlinearities},
   volume={107},
   ISSN={2469-9934},
   url={http://dx.doi.org/10.1103/PhysRevA.107.053713},
   DOI={10.1103/physreva.107.053713},
   number={5},
   journal={Physical Review A},
   publisher={American Physical Society (APS)},
   author={Shen, Yuan and Mok, Wai-Keong and Noh, Changsuk and Liu, Ai Qun and Kwek, Leong-Chuan and Fan, Weijun and Chia, Andy},
   year={2023},
   month=may }

@article{W_chtler_2023,
   title={Topological synchronization of quantum van der Pol oscillators},
   volume={5},
   ISSN={2643-1564},
   url={http://dx.doi.org/10.1103/PhysRevResearch.5.023021},
   DOI={10.1103/physrevresearch.5.023021},
   number={2},
   journal={Physical Review Research},
   publisher={American Physical Society (APS)},
   author={Wächtler, Christopher W. and Platero, Gloria},
   year={2023},
   month=apr }

@article{Murtadho_2023,
   title={Cooperation and Competition in Synchronous Open Quantum Systems},
   volume={131},
   ISSN={1079-7114},
   url={http://dx.doi.org/10.1103/PhysRevLett.131.030401},
   DOI={10.1103/physrevlett.131.030401},
   number={3},
   journal={Physical Review Letters},
   publisher={American Physical Society (APS)},
   author={Murtadho, Taufiq and Vinjanampathy, Sai and Thingna, Juzar},
   year={2023},
   month=jul }

@article{Solanki_2023,
   title={Symmetries and synchronization blockade},
   volume={108},
   ISSN={2469-9934},
   url={http://dx.doi.org/10.1103/PhysRevA.108.022216},
   DOI={10.1103/physreva.108.022216},
   number={2},
   journal={Physical Review A},
   publisher={American Physical Society (APS)},
   author={Solanki, Parvinder and Mehdi, Faraz Mohd and Hajdušek, Michal and Vinjanampathy, Sai},
   year={2023},
   month=aug }

@article{Cabot_2021,
   title={Synchronization and coalescence in a dissipative two-qubit system},
   volume={477},
   ISSN={1471-2946},
   url={http://dx.doi.org/10.1098/rspa.2020.0850},
   DOI={10.1098/rspa.2020.0850},
   number={2249},
   journal={Proceedings of the Royal Society A: Mathematical, Physical and Engineering Sciences},
   publisher={The Royal Society},
   author={Cabot, Albert and Luca Giorgi, Gian and Zambrini, Roberta},
   year={2021},
   month=may, pages={20200850} }

@article{Lee_2014,
   title={Entanglement tongue and quantum synchronization of disordered oscillators},
   volume={89},
   ISSN={1550-2376},
   url={http://dx.doi.org/10.1103/PhysRevE.89.022913},
   DOI={10.1103/physreve.89.022913},
   number={2},
   journal={Physical Review E},
   publisher={American Physical Society (APS)},
   author={Lee, Tony E. and Chan, Ching-Kit and Wang, Shenshen},
   year={2014},
   month=feb }

@article{Walter_2014,
   title={Quantum Synchronization of a Driven Self-Sustained Oscillator},
   volume={112},
   ISSN={1079-7114},
   url={http://dx.doi.org/10.1103/PhysRevLett.112.094102},
   DOI={10.1103/physrevlett.112.094102},
   number={9},
   journal={Physical Review Letters},
   publisher={American Physical Society (APS)},
   author={Walter, Stefan and Nunnenkamp, Andreas and Bruder, Christoph},
   year={2014},
   month=mar }

@article{W_chtler_2024,
   title={Topological Quantum Synchronization of Fractionalized Spins},
   volume={132},
   ISSN={1079-7114},
   url={http://dx.doi.org/10.1103/PhysRevLett.132.196601},
   DOI={10.1103/physrevlett.132.196601},
   number={19},
   journal={Physical Review Letters},
   publisher={American Physical Society (APS)},
   author={Wächtler, Christopher W. and Moore, Joel E.},
   year={2024},
   month=may }

@article{Laskar_2020,
   title={Observation of Quantum Phase Synchronization in Spin-1 Atoms},
   volume={125},
   ISSN={1079-7114},
   url={http://dx.doi.org/10.1103/PhysRevLett.125.013601},
   DOI={10.1103/physrevlett.125.013601},
   number={1},
   journal={Physical Review Letters},
   publisher={American Physical Society (APS)},
   author={Laskar, Arif Warsi and Adhikary, Pratik and Mondal, Suprodip and Katiyar, Parag and Vinjanampathy, Sai and Ghosh, Saikat},
   year={2020},
   month=jul }

@article{PhysRevResearch.5.033209,
  title = {Quantum synchronization of a single trapped-ion qubit},
  author = {Zhang, Liyun and Wang, Zhao and Wang, Yucheng and Zhang, Junhua and Wu, Zhigang and Jie, Jianwen and Lu, Yao},
  journal = {Phys. Rev. Res.},
  volume = {5},
  issue = {3},
  pages = {033209},
  numpages = {16},
  year = {2023},
  month = {Sep},
  publisher = {American Physical Society},
  doi = {10.1103/PhysRevResearch.5.033209},
  url = {https://link.aps.org/doi/10.1103/PhysRevResearch.5.033209}
}

@misc{li2025experimentalrealizationsynchronizationquantum,
      title={Experimental realization and synchronization of a quantum van der Pol oscillator},
      author={Yi Li and Zihan Xie and Xiaodong Yang and Yue Li and Xingyu Zhao and Xu Cheng and Xinhua Peng and Jun Li and Eric Lutz and Yiheng Lin and Jiangfeng Du},
      year={2025},
      eprint={2504.00751},
      archivePrefix={arXiv},
      primaryClass={quant-ph},
      url={https://arxiv.org/abs/2504.00751},
}

@article{Krithika_2022,
   title={Observation of quantum phase synchronization in a nuclear-spin system},
   volume={105},
   ISSN={2469-9934},
   url={http://dx.doi.org/10.1103/PhysRevA.105.062206},
   DOI={10.1103/physreva.105.062206},
   number={6},
   journal={Physical Review A},
   publisher={American Physical Society (APS)},
   author={Krithika, V. R. and Solanki, Parvinder and Vinjanampathy, Sai and Mahesh, T. S.},
   year={2022},
   month=jun }

@article{Koppenh_fer_2020,
   title={Quantum synchronization on the IBM Q system},
   volume={2},
   ISSN={2643-1564},
   url={http://dx.doi.org/10.1103/PhysRevResearch.2.023026},
   DOI={10.1103/physrevresearch.2.023026},
   number={2},
   journal={Physical Review Research},
   publisher={American Physical Society (APS)},
   author={Koppenhöfer, Martin and Bruder, Christoph and Roulet, Alexandre},
   year={2020},
   month=apr }

@misc{li2025twobodydissipatorengineeringenvironmentinduced,
      title={Two-body Dissipator Engineering: Environment-Induced Quantum Synchronization Transitions},
      author={Xingli Li and Yan Li and Yangqian Yan},
      year={2025},
      eprint={2506.07580},
      archivePrefix={arXiv},
      primaryClass={quant-ph},
      url={https://arxiv.org/abs/2506.07580},
}

@article{PhysRevA.108.032219,
  title = {Quantum effects on the synchronization dynamics of the Kuramoto model},
  author = {Delmonte, Anna and Romito, Alessandro and Santoro, Giuseppe E. and Fazio, Rosario},
  journal = {Phys. Rev. A},
  volume = {108},
  issue = {3},
  pages = {032219},
  numpages = {13},
  year = {2023},
  month = {Sep},
  publisher = {American Physical Society},
  doi = {10.1103/PhysRevA.108.032219},
  url = {https://link.aps.org/doi/10.1103/PhysRevA.108.032219}
}

@book{ROS01a,
  address = {Cambridge},
  author = {Rosenblum, M. G. and Pikovsky, A. and Kurths, J.},
  publisher = {Cambridge University Press},
  title = {Synchronization -- A universal concept in nonlinear sciences},
  year = 2001
}

@book{strogatz:2000,
  author = {Strogatz, Steven H.},
  publisher = {Westview Press},
  title = {Nonlinear Dynamics and Chaos: With Applications to Physics, Biology, Chemistry and Engineering},
  year = 2000
}

@article{PhysRevLett.64.821,
  title = {Synchronization in chaotic systems},
  author = {Pecora, Louis M. and Carroll, Thomas L.},
  journal = {Phys. Rev. Lett.},
  volume = {64},
  issue = {8},
  pages = {821--824},
  numpages = {0},
  year = {1990},
  month = {Feb},
  publisher = {American Physical Society},
  doi = {10.1103/PhysRevLett.64.821},
  url = {https://link.aps.org/doi/10.1103/PhysRevLett.64.821}
}

@article{pecora_carro,
author = {Pecora, Louis and Carroll, T.},
year = {1990},
month = {03},
pages = {821},
title = {Synchronization in chaotic system},
volume = {64},
isbn = {9780123968401},
journal = {Physical Review Letters},
doi = {10.1063/1.4917383}
}

@article{PhysRevLett.71.65,
  title = {Circuit implementation of synchronized chaos with applications to communications},
  author = {Cuomo, Kevin M. and Oppenheim, Alan V.},
  journal = {Phys. Rev. Lett.},
  volume = {71},
  issue = {1},
  pages = {65--68},
  numpages = {0},
  year = {1993},
  month = {Jul},
  publisher = {American Physical Society},
  doi = {10.1103/PhysRevLett.71.65},
  url = {https://link.aps.org/doi/10.1103/PhysRevLett.71.65}
}

@article{Sacha_2017,
   title={Time crystals: a review},
   volume={81},
   ISSN={1361-6633},
   url={http://dx.doi.org/10.1088/1361-6633/aa8b38},
   DOI={10.1088/1361-6633/aa8b38},
   number={1},
   journal={Reports on Progress in Physics},
   publisher={IOP Publishing},
   author={Sacha, Krzysztof and Zakrzewski, Jakub},
   year={2017},
   month=nov, pages={016401} }

@article{RevModPhys.95.031001,
  title = {Colloquium: Quantum and classical discrete time crystals},
  author = {Zaletel, Michael P. and Lukin, Mikhail and Monroe, Christopher and Nayak, Chetan and Wilczek, Frank and Yao, Norman Y.},
  journal = {Rev. Mod. Phys.},
  volume = {95},
  issue = {3},
  pages = {031001},
  numpages = {34},
  year = {2023},
  month = {Jul},
  publisher = {American Physical Society},
  doi = {10.1103/RevModPhys.95.031001},
  url = {https://link.aps.org/doi/10.1103/RevModPhys.95.031001}
}

@article{PhysRevLett.117.090402,
  title = {Floquet Time Crystals},
  author = {Else, Dominic V. and Bauer, Bela and Nayak, Chetan},
  journal = {Phys. Rev. Lett.},
  volume = {117},
  issue = {9},
  pages = {090402},
  numpages = {5},
  year = {2016},
  month = {Aug},
  publisher = {American Physical Society},
  doi = {10.1103/PhysRevLett.117.090402},
  url = {https://link.aps.org/doi/10.1103/PhysRevLett.117.090402}
}

@article{PhysRevLett.116.250401,
  title = {Phase Structure of Driven Quantum Systems},
  author = {Khemani, Vedika and Lazarides, Achilleas and Moessner, Roderich and Sondhi, S. L.},
  journal = {Phys. Rev. Lett.},
  volume = {116},
  issue = {25},
  pages = {250401},
  numpages = {6},
  year = {2016},
  month = {Jun},
  publisher = {American Physical Society},
  doi = {10.1103/PhysRevLett.116.250401},
  url = {https://link.aps.org/doi/10.1103/PhysRevLett.116.250401}
}

@article{PhysRevLett.121.035301,
  title = {Boundary Time Crystals},
  author = {Iemini, F. and Russomanno, A. and Keeling, J. and Schir\`o, M. and Dalmonte, M. and Fazio, R.},
  journal = {Phys. Rev. Lett.},
  volume = {121},
  issue = {3},
  pages = {035301},
  numpages = {6},
  year = {2018},
  month = {Jul},
  publisher = {American Physical Society},
  doi = {10.1103/PhysRevLett.121.035301},
  url = {https://link.aps.org/doi/10.1103/PhysRevLett.121.035301}
}

@article{PhysRevLett.132.183803,
  title = {Time Crystal in a Single-Mode Nonlinear Cavity},
  author = {Li, Yaohua and Wang, Chenyang and Tang, Yuanjiang and Liu, Yong-Chun},
  journal = {Phys. Rev. Lett.},
  volume = {132},
  issue = {18},
  pages = {183803},
  numpages = {7},
  year = {2024},
  month = {May},
  publisher = {American Physical Society},
  doi = {10.1103/PhysRevLett.132.183803},
  url = {https://link.aps.org/doi/10.1103/PhysRevLett.132.183803}
}

@article{PhysRevLett.128.080603,
  title = {Seeding Crystallization in Time},
  author = {Hajdu\ifmmode \check{s}\else \v{s}\fi{}ek, Michal and Solanki, Parvinder and Fazio, Rosario and Vinjanampathy, Sai},
  journal = {Phys. Rev. Lett.},
  volume = {128},
  issue = {8},
  pages = {080603},
  numpages = {6},
  year = {2022},
  month = {Feb},
  publisher = {American Physical Society},
  doi = {10.1103/PhysRevLett.128.080603},
  url = {https://link.aps.org/doi/10.1103/PhysRevLett.128.080603}
}

@article{PhysRevLett.111.234101,
  title = {Quantum Synchronization of Quantum van der Pol Oscillators with Trapped Ions},
  author = {Lee, Tony E. and Sadeghpour, H. R.},
  journal = {Phys. Rev. Lett.},
  volume = {111},
  issue = {23},
  pages = {234101},
  numpages = {5},
  year = {2013},
  month = {Dec},
  publisher = {American Physical Society},
  doi = {10.1103/PhysRevLett.111.234101},
  url = {https://link.aps.org/doi/10.1103/PhysRevLett.111.234101}
}

@article{PhysRevX.15.011010,
  title = {Nonreciprocal Synchronization of Active Quantum Spins},
  author = {Nadolny, Tobias and Bruder, Christoph and Brunelli, Matteo},
  journal = {Phys. Rev. X},
  volume = {15},
  issue = {1},
  pages = {011010},
  numpages = {21},
  year = {2025},
  month = {Jan},
  publisher = {American Physical Society},
  doi = {10.1103/PhysRevX.15.011010},
  url = {https://link.aps.org/doi/10.1103/PhysRevX.15.011010}
}

@article{Kuramoto:1975ebm,
    author = "Kuramoto, Yoshiki",
    editor = "Araki, Huzihiro",
    title = "{Self-entrainment of a population of coupled non-linear oscillators}",
    doi = "10.1007/BFb0013365",
    journal = "Lect. Notes Phys.",
    volume = "39",
    pages = "420--422",
    year = "1975"
}

@book{lyap,
    author={Pikovsky, A and Politi, A},
    title={Lyapunov Exponents: A Tool to Explore Complex Dynamics},
    publisher={Cambridge University Press},
    year={2016},
}

@article{PhysRevA.14.2338,
  title = {Kolmogorov entropy and numerical experiments},
  author = {Benettin, Giancarlo and Galgani, Luigi and Strelcyn, Jean-Marie},
  journal = {Phys. Rev. A},
  volume = {14},
  issue = {6},
  pages = {2338--2345},
  numpages = {0},
  year = {1976},
  month = {Dec},
  publisher = {American Physical Society},
  doi = {10.1103/PhysRevA.14.2338},
  url = {https://link.aps.org/doi/10.1103/PhysRevA.14.2338}
}

@book{Kuramoto2003,
  author = {Kuramoto, Yoshiki},
  isbn = {978-0-486-42881-9},
  note = {originally published: Springer Berlin, New York, Heidelberg, 1984},
  publisher = {Dover Publications},
  series = {Chemistry Series},
  title = {Chemical oscillations, waves, and turbulence},
  username = {rincedd},
  year = 2003
}

@article{Plenio,
  title = {The quantum-jump approach to dissipative dynamics in quantum optics},
  author = {Plenio, M. B. and Knight, P. L.},
  journal = {Rev. Mod. Phys.},
  volume = {70},
  issue = {1},
  pages = {101--144},
  numpages = {0},
  year = {1998},
  month = {Jan},
  publisher = {American Physical Society},
  doi = {10.1103/RevModPhys.70.101},
  url = {https://link.aps.org/doi/10.1103/RevModPhys.70.101}
}

@article{Daley2014,
   title={Quantum trajectories and open many-body quantum systems},
   volume={63},
   ISSN={1460-6976},
   url={http://dx.doi.org/10.1080/00018732.2014.933502},
   DOI={10.1080/00018732.2014.933502},
   number={2},
   journal={Adv. Phys.},
   publisher={Informa UK Limited},
   author={Daley, Andrew J.},
   year={2014},
   month={Mar},
   pages={77}
}

@misc{fazio2025manybodyopenquantumsystems,
      title={Many-Body Open Quantum Systems},
      author={Rosario Fazio and Jonathan Keeling and Leonardo Mazza and Marco Schirò},
      year={2025},
      eprint={2409.10300},
      archivePrefix={arXiv},
      primaryClass={quant-ph},
      url={https://arxiv.org/abs/2409.10300},
}

@inbook{Galve_2017,
   title={Quantum Correlations and Synchronization Measures},
   ISBN={9783319534121},
   ISSN={2364-9062},
   url={http://dx.doi.org/10.1007/978-3-319-53412-1_18},
   DOI={10.1007/978-3-319-53412-1_18},
   booktitle={Lectures on General Quantum Correlations and their Applications},
   publisher={Springer International Publishing},
   author={Galve, Fernando and Luca Giorgi, Gian and Zambrini, Roberta},
   year={2017},
   pages={393–420} }

@misc{rufo2025quantumsemiclassicalsignaturesdissipative,
      title={Quantum and Semi-Classical Signatures of Dissipative Chaos in the Steady State},
      author={Griffith Rufo and Sabrina Rufo and Pedro Ribeiro and Stefano Chesi},
      year={2025},
      eprint={2506.14961},
      archivePrefix={arXiv},
      primaryClass={cond-mat.stat-mech},
      url={https://arxiv.org/abs/2506.14961},
}

@article{Atas_2013,
   title={Distribution of the Ratio of Consecutive Level Spacings in Random Matrix Ensembles},
   volume={110},
   ISSN={1079-7114},
   url={http://dx.doi.org/10.1103/PhysRevLett.110.084101},
   DOI={10.1103/physrevlett.110.084101},
   number={8},
   journal={Physical Review Letters},
   publisher={American Physical Society (APS)},
   author={Atas, Y. Y. and Bogomolny, E. and Giraud, O. and Roux, G.},
   year={2013},
   month=feb }

@article{Delmonte_2025,
   title={Measurement-induced phase transitions in monitored infinite-range interacting systems},
   volume={7},
   ISSN={2643-1564},
   url={http://dx.doi.org/10.1103/PhysRevResearch.7.023082},
   DOI={10.1103/physrevresearch.7.023082},
   number={2},
   journal={Physical Review Research},
   publisher={American Physical Society (APS)},
   author={Delmonte, Anna and Li, Zejian and Passarelli, Gianluca and Song, Eric Yilun and Barberena, Diego and Rey, Ana Maria and Fazio, Rosario},
   year={2025},
   month=apr }

@article{ gda_EPJB,
        author = {Piccitto, Giulia and Rossini, Davide and Russomanno, Angelo},
        title = {The impact of different unravelings in a monitored system of free fermions},
        DOI= "10.1140/epjb/s10051-024-00725-0",
        url= "https://doi.org/10.1140/epjb/s10051-024-00725-0",
        journal = {Eur. Phys. J. B},
        year = 2024,
        volume = 97,
        number = 6,
        pages = "90",
}

@article{PhysRevB.108.094305,
  title = {Spatiotemporally ordered patterns in a chain of coupled dissipative kicked rotors},
  author = {Russomanno, Angelo},
  journal = {Phys. Rev. B},
  volume = {108},
  issue = {9},
  pages = {094305},
  numpages = {14},
  year = {2023},
  month = {Sep},
  publisher = {American Physical Society},
  doi = {10.1103/PhysRevB.108.094305},
  url = {https://link.aps.org/doi/10.1103/PhysRevB.108.094305}
}

@article{Passarelli_2025,
   title={Chaos and magic in the dissipative quantum kicked top},
   volume={9},
   ISSN={2521-327X},
   url={http://dx.doi.org/10.22331/q-2025-03-05-1653},
   DOI={10.22331/q-2025-03-05-1653},
   journal={Quantum},
   publisher={Verein zur Forderung des Open Access Publizierens in den Quantenwissenschaften},
   author={Passarelli, Gianluca and Lucignano, Procolo and Rossini, Davide and Russomanno, Angelo},
   year={2025},
   month=mar, pages={1653} }

@article{d7tm-9hkp,
  title = {Nonstabilizerness of a boundary time crystal},
  author = {Passarelli, Gianluca and Russomanno, Angelo and Lucignano, Procolo},
  journal = {Phys. Rev. A},
  volume = {111},
  issue = {6},
  pages = {062417},
  numpages = {14},
  year = {2025},
  month = {Jun},
  publisher = {American Physical Society},
  doi = {10.1103/d7tm-9hkp},
  url = {https://link.aps.org/doi/10.1103/d7tm-9hkp}
}

@article{PhysRevB.106.224308,
  title = {Dissipative time crystals with long-range Lindbladians},
  author = {Passarelli, Gianluca and Lucignano, Procolo and Fazio, Rosario and Russomanno, Angelo},
  journal = {Phys. Rev. B},
  volume = {106},
  issue = {22},
  pages = {224308},
  numpages = {13},
  year = {2022},
  month = {Dec},
  publisher = {American Physical Society},
  doi = {10.1103/PhysRevB.106.224308},
  url = {https://link.aps.org/doi/10.1103/PhysRevB.106.224308}
}

@misc{paulino2025thermodynamicscoupledtimecrystals,
      title={Thermodynamics of coupled time crystals with an application to energy storage}, 
      author={Paulo J. Paulino and Albert Cabot and Gabriele De Chiara and Mauro Antezza and Igor Lesanovsky and Federico Carollo},
      year={2025},
      eprint={2411.04836},
      archivePrefix={arXiv},
      primaryClass={quant-ph},
      url={https://arxiv.org/abs/2411.04836}, 
}

@article{PhysRevB.103.184308,
  title = {Boundary time crystals in collective $d$-level systems},
  author = {Prazeres, Luis Fernando dos and Souza, Leonardo da Silva and Iemini, Fernando},
  journal = {Phys. Rev. B},
  volume = {103},
  issue = {18},
  pages = {184308},
  numpages = {16},
  year = {2021},
  month = {May},
  publisher = {American Physical Society},
  doi = {10.1103/PhysRevB.103.184308},
  url = {https://link.aps.org/doi/10.1103/PhysRevB.103.184308}
}

@article{PhysRevResearch.7.013276,
  title = {Dissipative quantum chaos unveiled by stochastic quantum trajectories},
  author = {Ferrari, Filippo and Gravina, Luca and Eeltink, Debbie and Scarlino, Pasquale and Savona, Vincenzo and Minganti, Fabrizio},
  journal = {Phys. Rev. Res.},
  volume = {7},
  issue = {1},
  pages = {013276},
  numpages = {31},
  year = {2025},
  month = {Mar},
  publisher = {American Physical Society},
  doi = {10.1103/PhysRevResearch.7.013276},
  url = {https://link.aps.org/doi/10.1103/PhysRevResearch.7.013276}
}

\end{document}